\documentclass[letterpaper,12pt]{article} 
\usepackage{amsmath,amsfonts}

\providecommand{\e}{\mathrm{e}}
\providecommand{\bvec}[1]{\mathbf{#1}}
\providecommand{\dom}{{\partial\Omega}}
\providecommand{\sint}[1]{\int_\dom d^d{#1}\,}
\providecommand{\vint}[1]{\int_{\Omega} d^{d+1} {#1}\,}
\providecommand{\partdiff}[1]{\frac{\partial}{\partial#1}}

\title{Conformal Field Theory Correlators from Classical Scalar Field
Theory on $AdS_{d+1}$} 
\author{W.~M\"uck\thanks{e-mail:wmueck@sfu.ca}\ ~and
K.~S.~Viswanathan\thanks{e-mail:kviswana@sfu.ca}\\ \small Department
of Physics, Simon Fraser University, Burnaby, B.C., V5A 1S6 Canada}

\begin{document} 
\maketitle 
\begin{abstract} 
We use the correspondence between scalar field theory on $AdS_{d+1}$
and a conformal field theory on $\mathbb{R}^d$ to calculate the $3$-
and $4$-point functions of the latter. The classical scalar field
theory action is evaluated at treelevel.  
\end{abstract} 
\newpage

\renewcommand{\baselinestretch}{1.2}
\normalsize

\section{Introduction} 
\label{intro}
Since the suggestion of Maldacena about the
equivalence of the large $N$ limit of certain conformal field
theories in $d$ dimensions on one hand and supergravity on
$AdS_{d+1}$ on the other hand \cite{Maldacena}, theories on Anti de
Sitter spaces seem to have undergone a renaissance. After detailed
investigations in the past (see for example
\cite{Burges,Fronsdal,Burgess}), there has been a multitude of papers
relating to this subject in various aspects in the last months alone
(see \cite{Henningson} for a recent list of references). In
particular, the suggested correspondence was made more precise in
\cite{Gubser,Ferrara3,Witten}. According to these references one
identifies the partition function of the $AdS$ theory (with suitably
prescribed boundary conditions for the fields) with the generating
functional of the boundary conformal field theory. Thus, one has
schematically 
\begin{equation} 
\label{intro:equiv}
  Z_{AdS}[\phi_0] = \int_{\phi_0} \mathcal{D}\phi\, \exp(-I[\phi])
  \equiv Z_{CFT}[\phi_0] = \left\langle \exp\left(\sint{x} \,
  \mathcal{O} \phi_0]\right) \right\rangle.  
\end{equation}
The path integral on the l.h.s.\ is calculated under the restriction
that the field $\phi$ asymptotically approaches $\phi_0$ on the
boundary. On the other hand, the function $\phi_0$ is considered as a
current, which couples to the scalar density operator $\mathcal{O}$
in the boundary conformal field theory. Calculating the l.h.s.\ of
\eqref{intro:equiv} thus allows one to explicitely obtain correlation
functions of the boundary conformal field theory. Of course, since
the $2$- and $3$-point functions are fixed (up to a constant) by
conformal invariance \cite{Francesco}, one is especially interested
in calculating the cases $n>3$.

It is not only of pedagogical interest to consider the classical
approximation to the $AdS$ partition function, which is obtained by
inserting the solutions of the classical field equations into
$S[\phi]$. In fact, the suggested $AdS/CFT$ correspondence
\cite{Maldacena} involves classical supergravity on the $AdS$ side.
Moreover, it is instructive to study toy examples in order to better
understand this correspondence. A number of examples, including free
massive scalar and $U(1)$ gauge fields were studied in \cite{Witten}
and free fermions were considered in \cite{Henningson,Leigh}. Since a
free field theory will inevitably lead to a trivial (i.e.\ free)
boundary CFT, we feel it necessary to consider interactions. A short
note on interacting scalar fields is contained in \cite{Volovich}. In
this paper we will consider in detail a classical interacting scalar
field on $AdS_{d+1}$ at tree level. 

We recall here for convenience the formulae necessary for solving the
classical scalar field theory with Dirichlet boundary conditions. Let
us start with stating the action for a real scalar field in $d+1$
dimensions (Riemannian signature) with polynomial interaction,
\begin{equation} 
\label{intro:action}
  I[\phi] = \vint{x} \sqrt{g} \left[ \frac12 \left((\nabla \phi)^2
  +m^2
  \phi^2 \right) + \sum_{n\ge 3} \frac{\lambda_n}{n!} \phi^n \right].
\end{equation} 
The action \eqref{intro:action} yields the equation of motion 
\begin{equation} 
\label{intro:eqnmot}
  (\nabla^2 - m^2) \phi = \sum_{n\ge 3} \frac{\lambda_n}{(n-1)!}
  \phi^{n-1}.  
\end{equation} 
Using the covariant Green's function, which satisfies 
\begin{align} 
\label{intro:green}
  (\nabla^2 - m^2) G(x,y) &= \frac{\delta(x-y)}{\sqrt{g(x)}}\\
\intertext{and the boundary condition} 
\notag
  G(x,y)|_{x\in \dom} &= 0, 
\end{align} 
the classical field $\phi$ satisfying the equation of motion 
\eqref{intro:eqnmot} and a Dirichlet boundary condition on $\dom$ 
satisfies the integral equation
\begin{equation} 
\label{intro:phieqn}
  \phi(x) = \sint{y} \sqrt{h}\, n^\mu \partdiff{y^\mu} G(x,y) \phi(y)
  + \vint{y} \sqrt{g}\, G(x,y) \sum_{n\ge3} \frac{\lambda_n}{(n-1)!}
  \phi(y)^{n-1}, 
\end{equation} 
where $h$ is the determinant of the induced metric on $\dom$ and
$n^\mu$ the unit vector normal to $\dom$ and pointing outwards. A
perturbative expansion in the couplings $\lambda_n$ is obtained 
by using \eqref{intro:phieqn} recursively. We shall denote the
surface term in \eqref{intro:phieqn} by $\phi^{(0)}$ and the
remainder by $\phi^{(1)}$. Then, substituting the classical solution
\eqref{intro:phieqn} into \eqref{intro:action}, integrating by parts
and using the properties of the Green's function one obtains to tree
level
\begin{equation}
\label{intro:action2}
  I[\phi] = \frac12 \sint{x} \sqrt{h}\, n^\mu \phi^{(0)} \partial_\mu
  \phi^{(0)} + \sum_{n\ge 3} \frac{\lambda_n}{n!} \vint{x} \sqrt{g}\,
  \left(\phi^{(0)}\right)^n.
\end{equation} 

A short outline of the remainder of this paper is as follows. In
Sec.~\ref{field} we consider the free field on $AdS_{d+1}$. We
explicitely calculate the solutions to the wave equation, the Green's
function, solve the Dirichlet boundary problem and find the $2$-point
function of the boundary conformal field theory. In Sec.~\ref{tree}
we perform the calculations at tree level. An explicit closed formula
for the $n$-point function does not seem attainable for $n>3$.
However, we will stay general as far as possible and only then
specialize in the cases $n=3$ and $n=4$. Finally, Sec.~\ref{conc}
contains conclusions.

\section{Free Field Theory on $AdS_{d+1}$}
\label{field}
We will use the representation of $AdS_{d+1}$ as the upper half space 
$(x_0>0)$ with the metric 
\begin{equation} 
\label{field:metric}
  ds^2 = \frac{1}{x_0^2} \sum_{i=0}^{d} dx_i^2, 
  \end{equation} 
which possesses the constant curvature scalar $R=-d(d+1)$. The
boundary
$\dom$ is given by the space $\mathbb{R}^d$ with $x_0=0$ plus the
single point $x_0=\infty$ \cite{Witten}. In the sequel we shall adopt
the notations $x=(x_0,\bvec{x})$, $x^\ast=(-x_0,\bvec{x})$ and $x^2=
x_0^2 +\bvec{x}^2$.

Let us first solve the ``massive'' wave equation 
\begin{equation}
\label{field:wave}
  (\nabla^2-m^2)\phi = \left( x_0^2 \sum_{i=0}^d \partial^2_i -
  x_0(d-1) \partial_0 - m^2 \right) \phi = 0.  
\end{equation} 
The linearly independent solutions of \eqref{field:wave} are found to
be
\begin{equation} 
\label{field:modes}
  x_0^\frac{d}{2} \e^{-i\bvec{k\cdot x}} \begin{cases} I_\alpha(k
  x_0)\\ K_\alpha(k x_0) \end{cases}, \quad \text{where} \quad
  \alpha =\sqrt{\frac{d^2}{4}+m^2},
\end{equation} 
$\bvec{k}$ is a momentum $d$-vector and $k=|\bvec{k}|$. It is easy to
check that these modes are not square integrable, if $m^2\ge-d^2/4$.

The modes can now be used to calculate the Green's function on
\eqref{intro:green}. Making the ansatz 
\begin{equation}
\label{field:greenans}
  G(x,y) = \int \frac{d^d k}{(2\pi)^d}\, x_0^\frac{d}{2}
  \e^{-i\bvec{k\cdot}(\bvec{x}-\bvec{y})} f(k,y_0) \begin{cases}
  I_\alpha(k x_0) K_\alpha(k y_0) &\text{for $x_0<y_0$},\\
  K_\alpha(k x_0) I_\alpha(k y_0) &\text{for $x_0>y_0$},
  \end{cases} 
\end{equation} 
we explicitely satisfy the boundary condition at $x_0=0$ and $\infty$ 
and ensure continuity at $x_0=y_0$. Matching the two regions at the
discontinuity yields $f=-y_0^\frac{d}{2}$. The ansatz
\eqref{field:greenans} can be integrated and gives 
\begin{gather}
\label{field:green}
  G(x,y) = - \frac{c}{2\alpha} \xi^{-\Delta}
  F\left(\frac{d}{2},\Delta;\alpha+1;\frac{1}{\xi^2}\right),\\
\intertext{where $F$ denotes the hypergeometric function
\cite{Gradshteyn},} 
\label{field:xi}
  \xi = \frac{1}{2x_0y_0} \left[\frac12 \left((x-y)^2 +
  (x-y^\ast)^2\right) + \sqrt{(x-y)^2(x-y^\ast)^2} \right]
\end{gather}
and the new constants are defined by $\Delta=d/2+\alpha$ and
$c=\Gamma(\Delta)/(\pi^\frac{d}{2}\Gamma(\alpha))$. The Green's
function \eqref{field:green} coincides with the one found by Burgess
and L\"utken \cite{Burgess} after using a transformation formula for
the hypergeometric function \cite[formula 9.134 2.]{Gradshteyn}. Our
form has the advantage that for even $d$ the result can, using either
special value formulae or the definition as a series, be expressed in
terms of rational functions. For example, for $d=2$ we can use 
\[ F(1,1+\alpha;1+\alpha;z) = \frac{1}{1-z}. \]

We shall in this paper make use only of the boundary behaviour of the
Green's function. Since the induced metric diverges on the boundary
of $AdS_{d+1}$ ($x_0=0$), one has to consider the standard formalism
described in Sec.~\ref{intro} on a near-boundary surface
$x_0=\epsilon>0$ and then take the limit $\epsilon \to 0$. It has
been pointed out recently by Freedman \emph{et al.}\ \cite{Freedman}
that this limit has to be taken carefully, in particular at the 
very end of those calculations, which involve only the boundary
behaviour of the classical solution. It is therefore necessary to
find the Green's function, which vanishes not at $x_0=0$, but at
$x_0=\epsilon$. One can easily change \eqref{field:greenans} to
accomodate this. Denoting the new Green's function by $G_\epsilon$,
we find 
\begin{equation}
\label{field:greeneps}
  G_\epsilon(x,y) = G_0(x,y) + \int \frac{d^d k}{(2\pi)^d}\, 
  (x_0y_0)^\frac{d}{2} \e^{-i\bvec{k\cdot}(\bvec{x}-\bvec{y})}
  K_\alpha(k x_0) K_\alpha(k y_0) 
  \frac{I_\alpha(k\epsilon)}{K_\alpha(k \epsilon)},
\end{equation}
where $G_0$ is given by \eqref{field:greenans} and
\eqref{field:green}. It does not seem possible to perform the
momentum integral, but this is not necessary in order to obtain the
desired boundary behaviour. In particular, we find the normal
derivative on the boundary as 
\begin{equation}
\label{field:greendel}
  \left. \partdiff{y_0} G_\epsilon(x,y) \right|_{y_0=\epsilon} = -
  x_0^\frac{d}{2} \epsilon^{\frac{d}{2}-1} \int \frac{d^d
  k}{(2\pi)^d}\, \e^{-i\bvec{k\cdot}(\bvec{x}-\bvec{y})}
  \frac{K_\alpha(k x_0)}{K_\alpha(k \epsilon)},
\end{equation}
giving $-\epsilon^{d-1}\delta(\bvec{x}-\bvec{y})$ for $x_0=\epsilon$.

The bulk behaviour of the free field can be obtained from
\eqref{intro:phieqn} using the asymptotic behaviour of the Bessel
function in the denominator of \eqref{field:greendel} for
$\epsilon\to 0$. We note that for $AdS_{d+1}$ one has
$\sqrt{h(y)}=\epsilon^{-d}$ and $n^\mu =
(-\epsilon,\bvec{0})$. The minus sign comes from $n^\mu$ pointing
outward. One finds
\begin{equation}
\label{field:phibulk}
  \phi^{(0)bulk}(x) = c \epsilon^{\Delta-d} \int d^dy\,
  \phi_\epsilon(\bvec{y}) \left( \frac{x_0}{x_0^2 +
  |\bvec{x}-\bvec{y}|^2} \right)^\Delta,
\end{equation}
where $\phi_\epsilon$ denotes the Dirichlet boundary value at
$x_0=\epsilon$. We define
\begin{equation}
\label{field:phi0def}
  \phi_0(\bvec{x}) = \epsilon^{\Delta-d} \phi_\epsilon(\bvec{x})
\end{equation}
in order to make contact with the conformal field theory on the
boundary of $AdS_{d+1}$.

Eqn.~\eqref{field:phibulk} is the solution to the Dirichlet problem
with the boundary at $x_0=0$ \cite{Witten}. However, for the
two-point function we need to calculate the surface integral in
\eqref{intro:action2}, i.e.\ we need the near-boundary behaviour for
a boundary at $x_0=\epsilon$. Using the exact expression
\eqref{field:greendel} we find 
\begin{gather}
\notag
  \partial_0 \phi |_{x_0=\epsilon} = \frac1\epsilon \int d^dy\, 
  \phi_\epsilon(\bvec{y}) \int \frac{d^d k}{(2\pi)^d}\, 
  \e^{-i\bvec{k\cdot}(\bvec{x}-\bvec{y})} \left[ \frac{d}{2}-\alpha + 
  k \partdiff{k} \ln \left((k\epsilon)^\alpha K_\alpha(k\epsilon)
  \right) \right].\\
\intertext{The first two terms in the squared bracket yield $\delta$
function contact terms in the two-point funcion, which are of no
interest to us. In the third term, the divergence of the Bessel
function for $\epsilon\to0$ is exactly cancelled by the power of
$\epsilon$ in front of it. Using the series expansion}
\notag
  z^\alpha K_\alpha(z) = 2^{\alpha-1} \Gamma(\alpha) \left[ 1 -
  \frac{\Gamma(1-\alpha)}{\Gamma(1+\alpha)}
  \left(\frac{z}{2}\right)^{2\alpha} +\cdots \right],\\
\intertext{where the dots denote terms of order $z^n$ and
$z^{2\alpha+n}$, one can approximate the logarithm and then evaluate
the integral to obtain}
\label{field:phisurf}
  \partial_0 \phi |_{x_0=\epsilon} = 2 \alpha c \epsilon^{2\alpha-1}
  \int d^dy\,
  \frac{\phi_\epsilon(\bvec{y}}{|\bvec{x}-\bvec{y}|^{2\Delta}} +
  \cdots.
\end{gather}
Inserting \eqref{field:phisurf} into \eqref{intro:action2} we find
the value of the free field action as 
\begin{equation}
\label{field:action}
  I^{(0)} = - \frac12 \int d^dx d^dy\, 2\alpha
  c\epsilon^{2(\Delta-d)}
  \frac{\phi_\epsilon(\bvec{x})
  \phi_\epsilon(\bvec{y})}{|\bvec{x}-\bvec{y}|^{2\Delta}}
  +\cdots.
\end{equation}
Taking the limit $\epsilon\to0$ with the definition
\eqref{field:phi0def} we hence obtain, in agreement with
\cite{Freedman}, the two-point function for the boundary conformal
operators,
\begin{equation}
\label{field:2point}
   \langle \mathcal{O}(\bvec{x}) \mathcal{O}(\bvec{y}) \rangle =
   \frac{2\alpha c}{|\bvec{x}-\bvec{y}|^{2\Delta}}.
\end{equation} 

\section{Calculations at Tree Level} 
\label{tree}
At tree level, one can take the limit $\epsilon\to0$ beforehand,
which makes the considerations somewhat easier. The reason is that
the tree level calculations 
involve bulk integrals over $AdS_{d+1}$, as in the second term of 
\eqref{intro:action2}. Hence only the bulk behaviour of the free
field will 
be needed, which was obtained in Sec.~\ref{field}. Inserting
\eqref{field:phibulk} (with $\epsilon\to0$) into the interaction term
of the action 
one obtains
\begin{gather} 
\label{field:act1}
  I^{(1)}[\phi_0] = \sum_{n\ge3} \frac{c^n\lambda_n}{n!} \int
  d^dx_1\ldots d^dx_n\, \phi_0(\bvec{x}_1)\ldots \phi_0(\bvec{x}_n)
  I_n(\bvec{x}_1,\ldots,\bvec{x}_n),\\ \intertext{with}
\label{field:int}
  I_n(\bvec{x}_1,\ldots,\bvec{x}_n) = \int d^{d+1}y\,
  \frac{y_0^{-(d+1)+n\Delta}}{\left[
  \left(y_0^2+|\bvec{y}-\bvec{x}_1|^2\right) \ldots
  \left(y_0^2+|\bvec{y}-\bvec{x}_n|^2\right) \right]^\Delta}.
\end{gather} 
We can read off the connected part of the tree level $n$-point
functions $(n\ge3)$ for the operator $\mathcal{O}$ from
\eqref{field:act1}, 
\begin{equation} 
\label{field:npoint}
  \langle \mathcal{O}(\bvec{x}_1) \ldots \mathcal{O}(\bvec{x}_n)
  \rangle_\mathrm{conn.} = - \lambda_n c^n
  I_n(\bvec{x}_1,\ldots,\bvec{x}_n).  
\end{equation}

We shall now elaborate on a detailed calculation of the $3$- and
$4$-point functions. After a Feynman parametrization the $y$ integral
in \eqref{field:int} can be done yielding 
\begin{equation} 
\notag
  I_n = \frac{\pi^\frac{d}{2} \Gamma\left(\frac{n}{2}\Delta
  -\frac{d}{2}\right) \Gamma\left(\frac{n}{2}\Delta\right)}{%
  2\Gamma(\Delta)^n} \int\limits_0^\infty d\alpha_1 \ldots
  d\alpha_n\,
  \delta \left(\sum \alpha_i-1\right)
  \frac{\prod\alpha_i^{\Delta-1}}{\left(\sum\limits_{i<j}
  \alpha_i\alpha_j x_{ij}^2\right)^{\frac{n}{2}\Delta}}, 
\end{equation}
where $x_{ij}=|\bvec{x}_i-\bvec{x}_j|$. Now we can introduce new
integration variables $\beta_i$ by $\alpha_1=\beta_1$ and
$\alpha_i=\beta_1\beta_i$ $(i\ge2)$. The integration over $\beta_1$
is then trivial and leads to 
\begin{equation} 
\label{field:in}
  I_n = \frac{\pi^\frac{d}{2} \Gamma\left(\frac{n}{2}\Delta
  -\frac{d}{2}\right) \Gamma\left(\frac{n}{2}\Delta\right)}{%
  2\Gamma(\Delta)^n} \int\limits_0^\infty d\beta_2 \ldots d\beta_n\,
  \frac{\prod\limits_{i=2}^n \beta_i^{\Delta-1}}{\left[
  \sum\limits_{i=2}^n \beta_i \left(x_{1i}^2 +
  \sum\limits_{j>i}\beta_j
  x_{ij}^2\right)\right]^{\frac{n}{2}\Delta}}.  
\end{equation} 
We shall not try to perform the remaining integration in the general
formula, but consider the cases $n=3$ and $n=4$. For $n=3$ the
integrations can be carried out straightforwardly. Inserting the
result in \eqref{field:npoint} gives 
\begin{equation} 
\label{field:3point}
  \langle \mathcal{O}(\bvec{x}_1) \mathcal{O}(\bvec{x}_2)
  \mathcal{O}(\bvec{x}_3) \rangle =
  -\frac{\lambda_3\Gamma\left(\frac12\Delta+\alpha\right)}{2\pi^d}
  \left[\frac{\Gamma\left(\frac12\Delta\right)}{\Gamma(\alpha)}\right
  ]^3 \frac{1}{\left(x_{12} x_{13} x_{23}\right)^\Delta}.  
\end{equation}
For $n=3$ there is no disconnected contribution, hence
\eqref{field:3point} describes the full $3$-point function at tree
level. 

For $n=4$, we obtain after integration over $\beta_4$ and $\beta_3$ 
\[ I_4 = 
  \frac{\Gamma\left(2\Delta-\frac{d}{2}\right)}{2\Gamma(2\Delta)}
  \frac{\pi^\frac{d}{2}}{(x_{12}x_{34})^{2\Delta}}
  \int\limits_0^\infty
  \frac{d\beta_2}{\beta_2} \,
  F\left(\Delta,\Delta;2\Delta;1-\frac{(x_{13}^2+\beta_2
  x_{23}^2)(x_{14}^2+\beta_2 x_{24}^2)}{\beta_2 x_{12}^2
  x_{34}^2}\right). \] 
A change of integration variables and the introduction of the
conformal invariants (harmonic ratios) \cite{Francesco} 
\[  \beta_2=\frac{x_{13}x_{14}}{x_{23}x_{24}}\e^{2z},\qquad \eta =
  \frac{x_{12}x_{34}}{x_{14}x_{23}}, \qquad \zeta =
  \frac{x_{12}x_{34}}{x_{13}x_{24}}, \] 
then yields 
\begin{equation}
\label{field:4point}
  I_4= \frac{\Gamma\left(2\Delta-\frac{d}{2}\right)}{\Gamma(2\Delta)}
  \frac{2\pi^\frac{d}{2}}{\left(\eta\zeta\prod\limits_{i<j} x_{ij}
  \right)^{\frac23\Delta}} \int\limits_0^\infty dz\,
  F\left(\Delta,\Delta;2\Delta;1-\frac{(\eta+\zeta)^2}{(\eta\zeta)^2}
  -\frac{4}{\eta\zeta} \sinh^2z \right).  
\end{equation} 
Obviously, Eqn.~\eqref{field:4point} is of exactly the form dictated
for a four point function by conformal invariance \cite{Francesco}.

\section{Conclusions}
\label{conc}
We have considered an example of the correspondence between field
theories on an $AdS$ space and CFTs on its boundary. The classical
interacting scalar field has been treated at tree level and a
nontrivial conformal field theory of boundary operators has been
obtained. We calculated a non-trivial coefficcient of the $3$-point
function and, for the first time with this method, found an
expression for the function $f(\eta,\zeta)$ contained in the
$4$-point function \cite{Francesco}. We believe that the obtained
results will also be helpful for studying more complicated field
theories containing fermions and gauge fields.

\section*{Acknowledgements}
We are grateful to the authors of \cite{Freedman} for making us aware
of the subtleties of the $\epsilon\to0$ limit, which led to wrong
factors in all the correlators in an earlier version of this paper.

This work was supported in part by an operating grant from NSERC.
W.~M.\ gratefully acknowledges the support with a Graduate Fellowship
from Simon Fraser University.

\end{document}